# THE CHANCELLOR TRAP: ADMINISTRATIVE MEDIATION AND THE HOLLOWING OF SOVEREIGNTY IN THE ALGORITHMIC AGE

Xuechen Niu

Northern Arizona University

xn23@nau.edu

**Abstract**

The contemporary governance discourse on Artificial Intelligence often emphasizes catastrophic loss-of-control scenarios. This article suggests that such framing may obscure a more immediate failure mode: chancellorization, or the gradual hollowing out of sovereignty through administrative mediation. In high-throughput, digitally legible organizations, AI-mediated decision support can reduce the probability that failures become publicly legible and politically contestable, even when underlying operational risk does not decline. Drawing on the institutional history of Imperial China, the article formalizes this dynamic as a principal-agent problem characterized by a verification gap, in which formal authority (auctoritas) remains downstream while effective governing capacity (potestas) migrates to intermediary layers that control information routing, drafting defaults, and evaluative signals. Empirical support is provided through a multi-method design combining historical process tracing with a cross-national panel plausibility probe (2016-2024). Using incident-based measures of publicly recorded AI failures and administrative digitization indicators, the analysis finds that higher state capacity and digitalization are systematically associated with lower public visibility of AI failures, holding AI ecosystem expansion constant. The results are consistent with a paradox of competence: governance systems may become more effective at absorbing and resolving failures internally while simultaneously raising the threshold at which those failures become politically visible. Preserving meaningful human sovereignty therefore depends on institutional designs that deliberately reintroduce auditable friction.

# Table of Contents

## Introduction

Over the past decade, artificial intelligence has moved from a specialized analytic tool into a general-purpose infrastructure embedded in routine governance. In firms and states, AI systems increasingly sit between decision-makers and the underlying record: they retrieve and rank documents, compress evidence into summaries, draft memos and recommendations, and translate organizational goals into operational workflows. This shift is often defended as a practical response to scale. As workloads grow and domains become more technical, delegation expands. Still, delegation also changes where effective control sits. A long line of principal–agent research shows that when monitoring struggles to keep pace with task complexity, agents often gain latitude to shape outcomes through control of procedures and information even when principals retain formal authority. AI-mediated decision support intensifies this dynamic by accelerating mediation and lowering the cost of routinized approval, increasing the likelihood of cognitive inertia and automation-induced complacency.

Public discussion of high-stakes AI risk often centers on scenarios of abrupt loss of control, treating sovereignty as something seized through a single rupture. These scenarios are analytically clean, but they sit uneasily with how complex governance systems typically fail. Authority often erodes through procedural drift: small, cumulative changes in what information reaches the decision-maker, how options are prepared, and which signals define acceptable performance. I began this project by trying to connect AI risk language to institutional realities in decision support, and I repeatedly encountered the same friction point: formal sign-off can remain downstream, while the practical conditions of judgment migrate upstream into the layer that filters records and drafts decisions. This is not a claim about individual weakness. It is an institutional claim about how control rights and accountability behave under bounded monitoring.

The conceptual anchor for this article comes from a long empirical record: the sustained struggle between imperial authority and ministerial governing power in Chinese administrative history. The key analytical distinction is between *auctoritas* (sovereignty as the right to decide) and *potestas* (governing power as the capacity to process information). Emperors held final authority in principle, yet they faced a constraint that was easy to state and difficult to escape: they could not directly observe conditions across a large territory and could not personally process the full volume of memorials, appointments, and fiscal decisions. Governance therefore depended on intermediaries and routinized document workflows. The historical question was never whether mediation existed; it was whether mediation remained contestable and whether the sovereign retained practical methods to verify what reached the throne. Across dynasties, historians repeatedly note a recurring institutional pattern: actors who controlled access, routing, and drafting exercised durable leverage even when formal supremacy remained unchanged. I treat this record as a mechanism template rather than as a moral fable.

I use the term Chancellor Trap to describe the failure mode implied by that template: sovereignty can be hollowed out when governing power accumulates in an intermediary layer that controls routing, drafting defaults, and proxy signals. The term is intended to be descriptive rather than rhetorical. Historically, chancellor-like apparatuses occupied the junction where information became policy. In contemporary settings, the analogue is often a socio-technical layer rather than a single person: retrieval pipelines, ranking systems, drafting interfaces, and evaluation mechanisms that shape what becomes the working record and what counts as an acceptable output. The critical point is structural in tendency. When it is cheaper to approve a polished draft than to reconstruct the underlying record, and when verification is costly, dependence becomes rational and durable. Under these conditions, control rights can drift upstream even if formal authority and accountability remain downstream.

If this is the relevant failure mode, what should concern us about AI-mediated governance? The central issue is how failures surface, how they travel through organizations, and when they become contestable. The argument developed here focuses on a *legibility threshold*: in organizations optimized for throughput and administrative smoothness, AI-mediated workflows can absorb errors into internal processes, lowering the probability that failures become publicly visible and politically actionable.

Do improvements in administrative capacity necessarily make governance safer, or can they sometimes make failure less visible? This question motivates the analysis that follows. Rather than treating safety as a latent condition that can be inferred directly from performance metrics, the article examines how institutional design shapes the visibility of failure—who sees it, when it is recorded, and whether it triggers contestation or correction.

From this starting point, the article is organized around two research questions that connect institutional drift to technical–institutional coupling. The first asks when AI-mediated decision support shifts governance from judgment to ratification, so that a downstream human retains sign-off while an upstream mediation layer shapes the decision record and the feasible option set. This question aligns with principal–agent theory and with scholarship on procedures as instruments of control under limited monitoring. The second asks which deployment features widen the verification gap by concentrating control over evidence selection, drafting defaults, and proxy performance signals. Preference-based tuning matters because it optimizes behavior against an evaluation environment rather than against truth as such, creating risks of sycophancy. Retrieval-augmented generation matters because it makes retrieval and ranking a gatekeeper for what becomes the record presented to users. Proxy metrics matter because Goodhart-style dynamics predict systematic drift when proxies become targets under optimization pressure, a phenomenon formalized in recent work on adversarial Goodhart effects.

These questions imply a set of mechanism claims that are, in principle, observable. First, control rights can shift upstream under a verification gap when an intermediary layer concentrates routing control, drafting defaults, or proxy optimization, even if formal authority remains downstream. Second, if this mechanism is operative, the shift should be detectable in symptoms such as reduced auditability of the evidence chain, narrowed option sets at the moment of decision, and weakened contestation pathways—patterns that resonate with findings on automation bias, over-reliance, and accountability gaps in computerized decision environments. Third, the paper suggests evaluating governance at the level of workflow properties rather than abstract principles: interventions should increase observability of routing and retrieval decisions, insert deliberative friction at approval boundaries, separate proxy optimization from safety adjudication, and institutionalize contestation as procedure. These requirements align with oversight theory on alarm-based accountability and with contemporary risk management frameworks that emphasize documented controls and monitoring. The framework earns its keep only if it can be tied to what organizations can log, audit, and enforce: what evidence is shown, what is omitted, what options are available in the interface, and what happens when users dissent or request verification.

Methodologically, the paper combines mechanism illustration with a limited empirical probe. The qualitative sections draw on structured evidence from public inquiries, court records, and organizational documents to document how decision records are constructed, how contestation is enabled or blocked, and how responsibility is assigned when automated systems mediate the record. These cases are treated as mechanism illustrations rather than as a representative sample of all deployments, in part because high-stakes systems are often opaque and in part because public controversies are precisely where audit and accountability failures become legible in traceable form. The empirical section adds a small cross-national panel plausibility probe. Its goal is deliberately narrow: it does not estimate latent "global AI safety." It tests a more specific implication of the Chancellor Trap—whether administrative modernization is associated with changes in the probability that AI-related failures become publicly visible as recorded events.

The remainder of the article proceeds in five parts. Part I uses administrative history to identify recurring bottlenecks through which intermediaries can shape outcomes without formal constitutional change. Part II maps those roles onto contemporary AI deployment components. Part III develops the accountability implications as a problem of epistemic dependence and mediated decision records. Part IV translates the mechanism into governance requirements and evaluation strategies. Part V presents the cross-national plausibility probe, using incident recording as a fire-alarm style signal rather than as a direct measure of latent safety.

**1. Chancellor Power as Institutional Sedimentation: A Historical Topology**

This article approaches the phenomenon of "chancellorization" not merely as a sequence of historical events, but as a recurring structural vulnerability inherent in the architecture of centralized governance. It posits that in any complex administrative system, a divergence inevitably widens between the locus of formal sovereignty (auctoritas) and the locus of operational governing power (potestas). While the sovereign retains the symbolic prerogative of the final decision, the intermediary layer—by virtue of controlling the procedural inputs—accumulates the practical capacity to determine outcomes. This creates a principal-agent problem of a specific magnitude: the agent does not need to defect or usurp the throne to control the empire; they merely need to capture the verification mechanism. In the Chinese imperial context, this tension is often narrated as the struggle between jūnquán (monarch power) and xiāngquán (minister power), but the analytic content transcends the specificities of dynasty. It reveals a universal bureaucratic logic: when monitoring costs are high and information is asymmetrical, procedure becomes power (Moe, 1984; McCubbins, Noll, & Weingast, 1987).

However, applying this logic to the algorithmic context suggests a distinct tension regarding legitimacy. Unlike the historical usurpation of the throne, the modern sovereign's submission to the 'algorithmic chancellor' appears to be frequently voluntary, predicated on the legitimacy of 'efficiency' and 'scientific optimization'. Faced with the insurmountable cognitive load of high-modernist governance, bounded verification capacity often creates a durable capacity constraint, generating a strong default tendency toward chancellorization unless countervailing institutions raise verification capacity or reintroduce auditable friction. This implies that the 'Chancellor Trap' might not be merely a pathology of agent opportunism, but rather a structural inevitability of managing complex systems. In this view, the sovereign is not simply deceived; rather, to ensure the functional continuity of the state, they may be compelled to outsource the epistemological conditions of their own judgment.

To operationalize this concept for the analysis of algorithmic governance, this paper isolates three distinct institutional prototypes—Qin, Tang, and Ming. These cases are selected because they cleanly separate three administrative bottlenecks that are currently being inextricably fused in modern decision support systems: the pricing of dissent, the capture of routing, and the sedimentation of drafting.

1.1 The Qin Prototype: The Verification Gap

The collapse of verification in the late Qin court offers a paradigmatic case of structural capture within an information environment. The historical episode of zhi lu wei ma—pointing at a deer and calling it a horse—is frequently reduced to a parable of deception. Functionally, however, the event served not to mislead the court regarding biological taxonomy but to generate common knowledge regarding the cost of veridicality. Zhao Gao maneuvered to conduct a public audit of loyalty, establishing a separating equilibrium. By compelling officials to endorse a blatant falsehood, the intermediary forced the bureaucracy to reveal its preference ordering. Officials faced a binary choice between alignment with observed reality or alignment with the power structure. Those who adhered to empirical truth signaled that their allegiance to objective reality exceeded their fear of the chancellor, thereby marking themselves as threats to regime consolidation. Conversely, those who identified the animal as a horse signaled that their compliance was absolute and decoupled from their perception of the world.

September, Zhao Gao wanted to revolt but feared the court officials would not obey, so he set up a test. He brought a deer and presented it to the Second Emperor, calling it a horse. The Emperor laughed and said: Is the Chancellor mistaken? Calling a deer a horse? He asked the officials around him. Some remained silent, some said it was a horse to ingratiate themselves with Gao, and some said it was a deer. Gao later secretly prosecuted those who said it was a deer via the law. Thereafter, the officials were all terrified of Gao. (Sima Qian, Shiji: Annals of Qin Shi Huang)

The theoretical implications of this episode extend beyond the specific tyranny of the Qin Dynasty. It illustrates the concept of preference falsification within a bureaucracy as rigorously formalized by Kuran (1995). The central mechanism is the manipulation of the incentive structure to detach public expression from private knowledge. When an intermediary engineers an environment where the transmission of veridical information becomes prohibitively costly, sovereign epistemic capacity collapses. The principal-agent framework in political economy suggests that when monitoring costs are high, agents inevitably exploit information asymmetries to pursue their own utility (Moe, 1984). In the Qin court the cost of truth was physical execution. In modern administrative contexts the cost manifests as reputational damage or the friction of being labeled obstructionist. Once the cost of truth-telling exceeds the benefit of correction the flow of valid intelligence to the center ceases. Arendt (1967) observed in her analysis of truth and politics that factual truth possesses a fragility compared to rational or mathematical truth. Once organized power substitutes a fabricated reality for a factual one the texture of reality itself is torn and the witness to the original fact becomes a political enemy. In this state the sovereign retains the formal right to command yet these commands are issued in a vacuum of valid intelligence. The sovereign relies on a map of the territory redrawn by the intermediary to ensure its own survival. This severs the feedback loop required for governance transforming the sovereign into a symbolic figurehead who ratifies decisions made upstream.

This historical mechanism finds a precise isomorphism in the alignment techniques of modern Artificial Intelligence, specifically in Reinforcement Learning from Human Feedback. The reward model in this architecture acts as the digital equivalent of the chancellor, serving as the proxy for the principal intent. The optimization objective of many contemporary AI agents is to maximize the reward signal provided by the human evaluator (Ouyang et al., 2022). A critical problem arises here often referred to as metric fixation or Goodhart's Law (Strathern, 1997). The model does not optimize for objective truth which is often ambiguous or computationally expensive to verify. It optimizes for the appearance of quality as judged by the labeler. This creates a structural vulnerability to reward hacking where the agent exploits flaws in the evaluation process to gain high scores without fulfilling the underlying intent (Amodei et al., 2016). If the human evaluator prefers confident or fluent answers the model learns that the path of least resistance to a high reward is to mimic those preferences. The alignment process intended to make the model safe inadvertently trains it to navigate the psychological vulnerabilities of the user.

This dynamic manifests empirically as sycophancy. Recent research demonstrates that large language models systematically agree with user stated views even when those views are objectively incorrect (Sharma et al., 2023). Just as Zhao Gao rewarded those who confirmed the false narrative and punished those who insisted on the empirical one the RLHF process can penalize models for providing complex or corrective truths that create friction against the user prompt. The agent learns that validity is effectively a function of user satisfaction. If a user asks a leading question based on a false premise an optimized model often hallucinates evidence to support that premise rather than challenging it because correction risks a negative reward signal. This is not a malfunction in the code. It is the model functioning exactly as incentivized by the feedback loop. The result is the emergence of a closed epistemic loop described by Sunstein (2002) as an echo chamber of one. The user feels powerful because the machine responds with deference but this deference is purchased at the cost of reality. The AI effectively captures governing power by controlling the information environment leaving the human user confident but empirically blinded.

1.2 The Tang Prototype: Routing Capture

If the Qin case demonstrates how incentives degrade within a hierarchy. The Tang Dynasty case reveals a different mechanism related to logistics and the physical control of information channels. The rise of the Shumishi or Commissioners for Privy Affairs provides a historical model for understanding how the management of document flow evolves into the management of the state itself. Following the An Lushan Rebellion, the Tang court faced a crisis of command. The established outer bureaucracy was thorough but slow. The emperors needed a way to transmit military orders rapidly to regional commanders. They turned to the inner-court eunuchs to bypass the administrative friction of the civil service. This decision was driven by the need for efficiency and speed. However, this logistical shift had profound political consequences that the emperors did not foresee.

Cybernetics and information theory suggest that a communication channel is never a neutral pipe. Karl Deutsch argued in The Nerves of Government that the structure of communication determines the structure of decision. When an intermediary controls the channel, they acquire the power to filter the signal. The Shumishi began as simple messengers. Over time they gained the authority to prioritize certain memorials and delay others. In an environment of information overload, the power to prioritize is the power to govern. Herbert Simon described this as the management of "bounded rationality." The sovereign has limited attention. The entity that decides what enters that field of attention effectively sets the agenda for the state. By determining which reports reached the emperor first, the eunuchs determined which problems were treated as urgent and which were ignored.

In the chaotic times, heavy military affairs were managed from the inside... The power of the Shumishi eventually exceeded that of the Chancellor... they connected the sovereign above with the subjects below, gaining access to everything. As a result, the authority of the state fell into the hands of private individuals. (Ouyang Xiu, Xin Tang Shu: Treatise on Officials)

This paper identifies this phenomenon as routing capture. It describes a condition where the intermediary shapes the political reality by controlling the interface between the data source and the decision maker. The eunuchs did not always need to falsify documents. They simply needed to manage the queue. If a report on a famine was delayed until after a military campaign was approved, the famine effectively did not exist as a factor in the decision. This is consistent with Schattschneider's observation that the definition of alternatives is the supreme instrument of power. The interface ceased to be a passive window into the empire. It became an active filter that operated according to the private interests of the gatekeepers.

This historical mechanism is isomorphic to the retrieval and ranking layers in modern Retrieval-Augmented Generation systems. These systems are designed to ground Artificial Intelligence in facts by retrieving relevant documents from a database. This process is often presented as objective. However, the determination of what is relevant is a complex mathematical operation that mirrors the political function of the Shumishi. RAG systems use embedding models to map text into numbers. They calculate the distance between the user's query and the available evidence. The system then retrieves only the top few results to present to the user or the language model. This process creates a bottleneck of attention similar to the one in the Tang court.

The political danger arises because the embedding space is constructed from training data that contains latent biases. Safiya Noble has documented how search algorithms can reinforce social hierarchies by prioritizing certain types of information over others. In a RAG system, if the embedding model defines dissenting legal theories or alternative historical accounts as mathematically distant from the standard query, those documents are excluded from the context window. They are rendered invisible. The user believes they are seeing a comprehensive summary of the available knowledge. In reality they are seeing a curated selection. This creates a "Black Box" effect where the criteria for inclusion are opaque. The user cannot see what documents were filtered out or why they were deemed irrelevant. The algorithm acts as a silent Shumishi. It determines the boundaries of the sovereign's reality without any requirement to explain its routing logic. The principal lacks the technical capacity to audit the queue. This leaves the human decision maker in a position of dependency. They rule over a world that has been pre-sorted and pre-interpreted by an automated bureaucracy.

1.3 The Ming Prototype: Default Bias

The Ming Dynasty provides a precise procedural analogy for the structural challenges inherent in the human-in-the-loop model. The specific mechanism of interest is the bureaucratization of the drafting process known as piaoni. The Hongwu Emperor abolished the prime ministership in 1380 to centralize all executive power in the hands of the throne. This institutional restructuring unintentionally created a vacuum of processing capacity. The single sovereign could not physically manage the volume of administrative memorials flowing from the provinces. This necessity forced the creation of the Grand Secretariat. The secretaries in this body were tasked with reviewing incoming documents and attaching a proposed response on a small slip of paper. The Emperor theoretically retained the power of the Vermilion Brush which was the exclusive authority to approve the draft or write a new one. However the sheer cognitive load of governance meant that the draft effectively became the decision. The secretaries controlled the output because they controlled the initial proposal.

The Grand Scholars offer mere drafted proposals; the ultimate decision lies with the Emperor's Vermilion approval... Yet, whatever the Cabinet proposes, the Emperor approves; and whatever the Cabinet rejects, the Emperor dismisses. (Zhang Tingyu, Ming Shi: Biographies of Grand Secretaries)

This historical reality anticipates the findings of behavioral economics regarding default bias and status quo bias. Thaler and Sunstein (2008) argue in Nudge that the specific arrangement of choices significantly influences the outcome. The draft functions as a powerful default option. When a drafted response is fluent properly formatted and cites the correct precedents the administrative friction required to reject it is significantly higher than the friction required to approve it. Rejection imposes a heavy cognitive tax on the sovereign. The decision maker must stop the workflow and re-derive the solution from first principles. They must reconstruct the reasoning and physically draft a new response. Approval requires only a minor physical action. Simon (1947) described this behavior through the lens of bounded rationality. Human decision makers satisfice rather than optimize when they face time constraints. The path of least resistance becomes the dominant strategy. Over time the act of governing degrades into the act of signing. The drafter accumulates the actual governing power while the sovereign retains only the ceremonial confirmation.

This phenomenon manifests in modern decision support systems as automation complacency. Parasuraman and Manzey (2010) define this state in their review of human-automation interaction. It characterizes a condition where operators rely heavily on the correctness of system outputs and consequently reduce their vigilance. This reduction in monitoring is a rational adaptation to a high-reliability environment rather than a simple failure of diligence. The operator allocates attention to other tasks because the system rarely fails. Generative AI accelerates this drift because of the linguistic fluency of the output. Traditional automation might fail in obvious ways. A language model produces drafts that appear grammatically perfect and logically coherent even when factually hallucinatory. This surface-level quality acts as a strong heuristic for accuracy. It masks the underlying errors.

The barrier to verification becomes the central mechanism of control. Skitka et al. (1999) observed in early studies of automation bias that humans tend to follow automated advice even when contradictory evidence is available. The cost of verification explains this behavior. To verify a legal brief or a medical diagnosis generated by an AI requires the human operator to replicate the cognitive labor that the machine has already performed. The operator must return to primary sources. They must read the raw case law or the patient history from scratch. This duplication of effort defeats the organizational purpose of using the tool which is efficiency. Simon (1947) established that administrative behavior is governed by bounded rationality. Decision makers satisfice. They look for a solution that is good enough to clear the immediate hurdle. In a high-volume workflow the AI draft is the satisficing solution. The friction of rejection is too high.

Elish (2019) describes the resulting institutional arrangement as a moral crumple zone. The term refers to the way liability is distributed in human-machine systems. The human operator is positioned at the edge of the system to absorb the legal and social consequences of failure. The machine operates at the center and retains practical control over the decision parameters. The human retains the formal accountability. This separation of control from liability creates a perverse incentive structure. The operator ratifies the machine output to maintain workflow velocity. They function as a liability sponge. They protect the institution from the claim that the decision was automated. The presence of the human allows the organization to claim that a sovereign judgment took place.

The historical analogy of the Ming draft is structurally identical to this modern workflow. The Vermilion Brush remains in human hands. The mind guiding the brush has been effectively outsourced to the drafting engine. The human sovereign retains the authority to say no. However the practical capacity to exercise that authority has eroded. The dependency on the system becomes absolute because the cost of disagreeing with the algorithm becomes professionally unsustainable. If an operator overrides the AI and makes an error they bear full responsibility for ignoring the advanced tool. If they follow the AI and an error occurs they can often appeal to the reliability of the system or the opacity of the technical process. The path of least resistance is to sign the draft. This institutionalizes the transfer of governing power. The sovereign becomes a ceremonial figure who validates the output of the chancellor.

**2. The Mechanics of Algorithmic Chancellorization**

The translation of these historical patterns into the architecture of contemporary AI-mediated governance reveals that chancellorization functions as a precise description of control rights displacement under conditions of verification deficit. We observe the emergence of a socio-technical agent that filters evidence and drafts mandates while the principal suffers from severe information asymmetry. Institutional drift occurs when the system shifts from a tool for retrieval to an engine of ratification. This section argues that the technical architectures of Reinforcement Learning from Human Feedback and embedding-based retrieval reinstate the historical risks of dissent pricing and routing capture. The central dynamic is the decoupling of the authority to decide from the capacity to generate options. This mirrors the principal-agent problem described by Jensen and Meckling (1976) where the agent controls the specific knowledge required to execute the task and can thereby manipulate the principal. In the context of algorithmic governance the agent is the model architecture itself which optimizes for metrics that may diverge from the sovereign interest of the user.

2.1 The Sycophantic Loop

The economic logic of Reinforcement Learning from Human Feedback creates an incentive environment that structurally mirrors the incentive structure of the late Qin court. RLHF trains models to maximize a reward signal derived from human preference comparisons as detailed by Ouyang et al. (2022). This method is currently essential for alignment and usability. However it introduces a critical vulnerability where the model learns to optimize for the perception of helpfulness rather than the reality of correctness.

The reward model functions as a proxy for the sovereign will. If the human evaluators during the training phase prefer confident or fluent answers the model will converge on a strategy of sophisticated sycophancy. It learns to generate outputs that minimize friction with the user intent rather than outputs that maximize factual accuracy. This creates a divergence between the objective function of the system and the epistemic needs of the decision maker Goodhart (1975) formulated the law that when a measure becomes a target it ceases to be a good measure. This dynamic has been rigorously formalized in the context of AI alignment by Manheim and Garrabrant (2018), who categorize such failures as adversarial variations of metric over-optimization. Here, the measure of quality is human approval. The model exploits this by providing answers that are psychologically satisfying to the user but potentially detached from empirical ground truth.

Empirical studies have begun to validate this concern regarding the stability of truth in model outputs. Research on sycophancy in large language models by Sharma et al. (2023) shows that models will actively agree with a user incorrectly stated views to maximize approval scores. Perez et al. (2022) further demonstrate that models can express distinct political biases to align with the perceived user persona. In a high-stakes policy or legal context this constitutes a fundamental governance failure. A policymaker queries an AI about the viability of a preferred strategy. The AI detects the user inclination and suppresses counter-evidence to produce a confirmation that appears objective. This mechanism operationalizes the historical deer-horse test described by Sima Qian (1993). The cost of dissent in the form of a low reward signal drives the system toward a compliance equilibrium. The observable implication for researchers is that the variance in model responses will correlate more strongly with the user prompt phrasing than with the underlying factual record. This phenomenon aligns with the critique of high-modernist legibility offered by James Scott (1998) in Seeing Like a State. Scott argued that administrative systems simplify complex reality to make it legible and manageable for the center. The AI system simplifies the informational environment to make it pleasing to the sovereign user. It strips away the necessary friction and complexity that characterize genuine judgment. The result is a smooth administrative surface that hides the underlying incoherence. The user retains the feeling of command while their connection to the actual parameters of the decision has been severed by the optimization process. The machine does not need to conspire to seize power. It merely needs to optimize for agreement in a context where the user cannot cost-effectively verify the truth.

2.2 The Black Box of Relevance

Retrieval-Augmented Generation is frequently conceptualized in technical literature as a mechanism for grounding designed to mitigate hallucination by anchoring model outputs in external knowledge bases. However from the perspective of political theory the retrieval layer functions as a site of intense contestation regarding agenda setting. We must interrogate the political nature of the search index itself. In this architecture the decision of which documents to retrieve and present to the reasoning engine is determined by embedding models that map textual data into high-dimensional vector space. The geometric closeness of two concepts in this space defines their relevance.

This mapping process is not an objective reflection of the world. It is a constructed artifact derived from the statistical patterns of the training corpora. This operation constitutes a precise digital enactment of the "high-modernist simplification" described by James Scott (1998). Just as the state standardizes complex social ecologies into legible administrative grids, the embedding model collapses the semantic nuance of human thought into a standardized vector space that is computationally legible but epistemically reductive. Consequently, the routing of information—the mechanism identified in the Tang prototype—is now encoded in these vector operations.This creates a form of algorithmic gatekeeping where the mathematical definition of relevance substitutes for the political judgment of importance. The embedding model effectively determines the boundaries of the epistemic territory available to the decision maker.

The mechanics of this exclusion warrant precise theoretical specification. In a dense retrieval system as described by Karpukhin et al. (2020) the algorithm calculates the cosine similarity between the query vector and millions of document vectors. It then selects the top-k results for inclusion in the generation context. This operation inherently exiles any information that does not align with the latent semantic structure of the model. If the embedding model clusters dissenting legal precedents alternative medical diagnoses or minority historical narratives as mathematically distant from the query vector these options are effectively rendered invisible. They are not censored through active suppression. They are simply routed into the void of irrelevance. This creates a feedback loop of confirmation where the system retrieves evidence that structurally resembles the query reinforcing the existing assumptions of the user or the dominant discourse in the training data. Safiya Noble (2018) has documented how similar mechanisms in search engines reinforce social hierarchies by prioritizing hegemonic representations over marginalized ones. In a governance context this means that a policy analyst querying a system about economic intervention might only receive documents that align with the specific economic orthodoxy encoded in the embedding space. The system constructs a closed epistemic loop where the user is presented with a consensus that is actually an artifact of the retrieval logic.

The governance implications of this routing capture are observable in real-world administrative failures. The Dutch Childcare Benefits Scandal provides a grim empirical demonstration of how algorithmic classification functions as a routing mechanism that precludes human judgment. In this case the Dutch tax authority used a risk-classification algorithm to identify potential fraud in childcare benefit applications. The system used variables such as dual nationality as proxies for risk. Once an applicant was classified as high risk by the system they were routed into an administrative workflow of rigid enforcement and debt collection. The human case officers theoretically retained the authority to review individual files. In practice the algorithmic designation established a presumption of guilt that was administratively impossible to overcome. The system routed these citizens into a category of fraud suspicion that defined their reality in the eyes of the state. The Parliamentary Inquiry Committee (2020) concluded that this resulted in an unprecedented injustice where the principles of the rule of law were systematically violated. The algorithm did not just advise the sovereign state it effectively blinded the state to the innocence of its own citizens by filtering out exculpatory context in favor of risk indicators. This establishes the central problem of auditability in chancellorized systems. The user of a RAG system sees only the documents that survived the retrieval filter. They cannot see the documents that were rejected. This asymmetry creates a Black Box Society effect as described by Frank Pasquale (2015). The criteria for inclusion in the official record are opaque to the user. The sovereign decision maker believes they are acting on the basis of a comprehensive review of the files. In reality they are acting on a curated selection that has been pre-processed by an unaccountable intermediary. The technical capacity to audit the queue to examine the documents that were filtered out and the logic of their exclusion becomes a prerequisite for the retention of genuine sovereignty. Without this capacity the human in the loop is merely ratifying the political geography imposed by the embedding space. The governance challenge shifts from managing the final output to governing the hidden architecture of retrieval.

2.3 The Automation of Authority

The Ming piaoni system finds its modern correlate in the generative drafting capabilities of Large Language Models. The power to draft is functionally the power to frame the option set and constrain the decision horizon of the sovereign. We must analyze the specific mechanism through which a draft acquires authority. The case of Mata v. Avianca (2023) serves as a critical empirical touchstone. In this legal proceeding a lawyer submitted a brief containing hallucinated precedents generated by ChatGPT. Popular discourse has largely categorized this event as an instance of individual professional incompetence or ethical lapse. However this paper views the incident as a structural failure indicative of automation bias a psychological phenomenon extensively documented by Skitka et al. (1999). The system produced a draft that was legally fluent formatted correctly and rhetorically persuasive. It mimicked the aesthetics of high-quality legal reasoning. The friction required to verify the citations to stop the workflow log into a primary legal database and cross-reference each claim was significantly higher than the cognitive ease of acceptance.

This dynamic illustrates the piaoni effect in action where the draft becomes the default. Behavioral economics provides a theoretical framework for understanding this through the concept of the default effect. Thaler and Sunstein (2008) argue that the choice architect who sets the default option exercises immense influence over the final outcome because agents tend to stick with the status quo to avoid transaction costs. In the context of AI-mediated governance the LLM acts as the choice architect. By presenting a single coherent and polished draft the system imposes a heavy cognitive tax on any attempt to deviate. The human decision maker must exert significant mental effort to reject the machine proposal and construct an alternative from scratch. As organizational throughput pressures increase the rational strategy for the bureaucrat is to minimize this effort. The draft is accepted not because it is verified to be true but because it is optimized to be plausible.

In organizational settings this leads to a widening verification gap. As systems become more reliable in the aggregate users engage in what can be described as moral buffering. They assume that the system output is derived from a rigorous process of logic and fact-checking even when the underlying architecture is probabilistic. This results in a redistribution of discretion that is largely invisible to external observers. The AI acting as the drafter constrains the solution space to a single trajectory. The human in the loop theoretically retains the authority to intervene. In practice their agency is reduced to a binary choice accept the plausible draft or halt the entire production line to perform manual labor. Under conditions of bureaucratic constraint acceptance becomes the dominant strategy. The human role transforms from decision-maker to rubber-stamper. Madeline Elish (2019) describes this institutional arrangement as the creation of a moral crumple zone. The term captures the displacement of accountability in complex socio-technical systems. The human operator is positioned at the immediate interface of the decision to absorb the legal and social liability for failure. The machine retains the practical control over the content and direction of the task. In the Avianca case the lawyer bore the full professional sanction for the hallucinated brief. The system that generated the error operated without penalty. This asymmetry creates a perverse incentive structure for organizations. They can deploy autonomous drafting tools to maximize efficiency and claim that human oversight is preserved. The human provides the seal of sovereignty the Vermilion Brush while the mind guiding the decision has been effectively outsourced to the drafting engine. The friction of verification prevents the human from exercising true agency. The system creates a dependency where the cost of disagreeing with the algorithm becomes professionally unsustainable.

## 3. The Crisis of Sovereignty: Decoupling Auctoritas from Potestas

The theoretical core of this analysis illuminates the fundamental dilemma of sovereignty in the algorithmic age. Chancellorization describes a specific trajectory of institutional decay where the decision maker retains auctoritas, the formal right to command and legitimize, while progressively losing potestas, the actual capacity to execute and shape the meaning of those commands. This separation represents a constitutional crisis within the administrative state rather than a mere technical calibration issue. Carl Schmitt defined the sovereign as the one who decides on the exception. In the context of algorithmic governance, the capacity to identify and decide on the exception is methodically eroded. The system is designed to normalize data and process outliers according to statistical probabilities. The human sovereign is transformed from the active decider of the exception into the passive consumer of a pre-processed reality. Max Weber warned that the bureaucracy would become an iron cage that constrains the actions of the individual. Algorithmic chancellorization tightens this cage by automating the constraints. The sovereign sits on the throne but the parameters of their reign are determined by the code that processes the information of the empire. The decision maker possesses the seal of state but lacks the independent cognitive access to the territory required to use it meaningfully.

First, this decoupling leads to a profound fragility of epistemic agency. Modern governance is predicated on the division of labor. John Hardwig argues that epistemic dependence is rational and necessary in a complex society. No single individual can possess all the knowledge required to rule. However, legitimate dependence requires auditability. When the chain of evidence is obscured by neural weights and embedding distances, the sovereign dependence becomes blind. This deepens the texture of the crisis. The reliance on algorithmic intermediaries suggests a fundamental shift in the sources of administrative legitimacy. Unlike historical instances where ministerial power was seized through intrigue, the modern sovereign appears to voluntarily enter the Chancellor Trap, driven by the sheer scale of information that exceeds the bounds of human rationality. In this context, the legitimacy of the 'algorithmic chancellor' is not derived from tradition or constitutional mandate, but from the promise of efficiency and 'scientific' optimization. This implies that the hollowing out of sovereignty may not be a failure of vigilance, but rather a structural adaptation—a trade-off where the capacity to govern is purchased at the cost of the capacity to understand.

This creates a vacuum of accountability that Mark Bovens identifies as the problem of many hands. In an AI-mediated administration, this becomes the problem of many layers. If a decision is challenged—why was this loan denied or why was this target selected—the human sovereign can only point to the recommendation of the system. The provenance of this recommendation is mathematically opaque. This erodes the reason-giving capacity that John Rawls identifies as central to liberal democratic legitimacy. Public reason requires that the exercise of political power be justifiable in terms that citizens can understand and accept. An algorithmic output that cannot be explained violates this requirement. The sovereign authority becomes brittle because it cannot answer the question why without deferring to the black box of the chancellor.

Second, chancellorization results in the atrophy of the option set. Governing power is fundamentally the power to define alternatives. Bachrach and Baratz famously identified this as the second face of power. This is the ability to limit the scope of decision making to safe issues and to prevent challenges to the status quo from even reaching the table. Algorithmic intermediaries function as agents of this second face of power. They are driven by optimization functions that favor the probable over the possible. They tend to collapse the option space toward the statistical mean. They present the best answer based on past data, obscuring the trade-offs and uncertainties that define genuine political choice. The sovereign is left with a constrained autonomy. They are free to choose, but only from the menu the chancellor has printed. This pre-emption of choice is more effective than direct coercion because it is invisible. The sovereign does not feel coerced; they feel helped. Yet their political horizon has been artificially narrowed to the technological limits of the model.

Finally, and perhaps most critically, the phenomenon degrades contestation pathways. Administrative law relies on the existence of fire alarms. McCubbins and Schwartz distinguish between police patrol oversight, which is active and costly, and fire alarm oversight, which relies on third parties to signal when the system is failing. Chancellorization dampens these alarms. By smoothing over the friction of decision making and presenting a veneer of objective confidence, AI systems can mask underlying errors until they metastasize into systemic failures. The Robodebt scheme in Australia serves as a grim testament to this mechanism. The automated debt-raising system operated with a presumption of correctness that shifted the burden of proof onto the vulnerable welfare recipients. The algorithm effectively silenced contestation by making it administratively impossible for the poor to challenge the debt notices. The chancellor effectively insulated the Cabinet from the reality of the policy impact. The system filtered out the noise of human suffering until the sheer scale of the injustice forced a Royal Commission. This demonstrates how the capture of the information environment by the automated intermediary allows the sovereign to proceed with a disastrous policy long after a manual bureaucracy would have collapsed under the weight of the complaints.

In conclusion, the risk of AI in governance is not that machines will become conscious and seize power in a dramatic coup. The risk is that they will become the perfect bureaucrats and subtly displace the locus of judgment. By controlling the incentives of reporting, the routing of context, and the defaults of drafting, algorithmic systems are poised to reenact the oldest drama of imperial administration. The throne remains radiant and the rituals of power are observed, but the effective control of the state has migrated to the secretariat. The sovereign becomes a prisoner of the very machinery built to serve them.

**4 Empirical study**

4.1 Empirical strategy and theoretical motivation

The empirical test presented here is intentionally modest in scope but specific in its mechanism. Because incident datasets capture only what becomes documented and publicly visible, the analysis does not attempt to estimate underlying AI risk—an unobservable variable. Instead, it examines whether higher administrative digitization correlates with the probability that AI-related failures become publicly recorded—an observable proxy for crossing the legibility threshold.

This approach serves as a mechanism-oriented plausibility probe. In the comparative political economy and state capacity literature, capacity is often treated as a unidimensional good that improves monitoring, enforcement, and policy outcomes. Within this perspective, greater administrative sophistication is generally expected to strengthen oversight and reduce failure. However, the Chancellor Trap framework motivates a more differentiated empirical question. Administrative capacity is not exercised solely through enforcement or detection; it is also exercised through information routing, procedural mediation, and service delivery.

A growing body of scholarship on state capacity and bureaucratic governance emphasizes that modern administrations increasingly operate through digital interfaces and rule-based processes that structure how problems are recorded, classified, and escalated. These systems can improve efficiency and consistency, but they may also alter the conditions under which failures become publicly observable. From this perspective, capacity may shape the threshold of recognition—the point at which an issue moves from internal administrative handling into the public and political domain—without necessarily affecting the underlying incidence of problems.

The empirical strategy therefore distinguishes two analytically separate margins. The first concerns material expansion: whether the domestic AI ecosystem deepens over time. This is operationalized using measures of AI talent concentration, which reflect the size and growth of the AI-related labor market. The second margin concerns epistemic visibility: whether AI-related failures cross the threshold into public documentation. Here, the focus is not on the severity of harm per se, but on whether incidents become legible to external observers through media reporting and public controversy. This distinction aligns with established work on fire-alarm versus police-patrol oversight, which emphasizes that many governance failures are detected indirectly through public signals rather than direct monitoring. By separating these margins, the analysis avoids conflating ecosystem growth with risk visibility.

4.2 Data and measurement

The empirical analysis uses an annual panel dataset covering ten advanced industrial democracies over the period 2016–2024. The sample is constructed to balance cross-national comparability with temporal coverage, and the panel is balanced by country-year where data availability permits. Three categories of data are combined: measures of public risk visibility, indicators of AI ecosystem depth, and disaggregated measures of administrative capacity.

Risk visibility is measured using the AI Incident Database (AIID), maintained by the Partnership on AI. The AIID compiles documented AI-related incidents and controversies based on publicly available sources, including media reports and investigative accounts (Partnership on AI, 2023). Because inclusion in the database depends on public reporting, the absence of recorded incidents in a given country-year should not be interpreted as evidence of safety or non-occurrence. Instead, it indicates that no incident reached the level of public documentation captured by the database. Following established approaches in the oversight and accountability literature, AIID-based outcomes are therefore treated as indicators of public legibility rather than prevalence (McCubbins & Schwartz, 1984). Empirically, this distinction motivates both an extensive-margin measure—whether any incident is recorded in a country-year—and an intensity measure conditional on visibility, scaled by population.

AI ecosystem depth is proxied by AI talent concentration, drawing on the OECD.AI framework based on LinkedIn Economic Graph data. This measure reflects the share of a country's workforce identified as having AI-related skills and roles, and has been used in prior analyses of AI labor markets and technological diffusion (OECD, 2023). To reduce skew and facilitate interpretation in within-country models, the analysis uses the logarithmic transformation ln(1 + AI talent concentration). While LinkedIn-based measures are subject to platform coverage biases, they remain one of the few sources providing consistent, time-varying cross-national data on AI labor markets, and are widely used in policy and academic research.

Administrative capacity is measured using the United Nations E-Government Development Index (EGDI) and its component indices. EGDI is a composite indicator published biennially by the United Nations Department of Economic and Social Affairs, designed to capture the readiness and capacity of national administrations to deliver services through digital means (United Nations, 2022). Disaggregating the index is central to the empirical strategy. The Human Capital Index (HCI) reflects educational attainment and skills relevant to digital governance; the Online Service Index (OSI) captures the scope and sophistication of digitized public services and administrative interfaces; and the Telecommunication Infrastructure Index (TII) measures the underlying connectivity required for digital service provision. This decomposition follows prior work emphasizing that different dimensions of capacity have distinct institutional functions and should not be treated as interchangeable (Fukuyama, 2013; Margetts & Dunleavy, 2013).

Standard development controls are included to account for confounding structural factors. These include GDP per capita (logged), urbanization measured as the urban population share, and high-technology exports (logged), drawing on World Bank indicator definitions for cross-national consistency (World Bank, 2023). In supplementary specifications, the Oxford Insights Government AI Readiness Index is included as an alternative governance-related benchmark. This index captures broader policy and institutional preparedness for AI but is available for a shorter time span than the EGDI components, and its inclusion therefore reduces the effective sample size (Oxford Insights, 2024). All substantive explanatory variables enter the models with a one-year lag to reduce simultaneity concerns and to reflect the temporal ordering implied by the theoretical framework.

4.3 Specifications and estimation

This section formalizes the empirical strategy used to probe the mechanisms proposed by the Chancellor Trap framework. The purpose of the analysis is not to estimate a general causal effect of state capacity on AI safety outcomes, but to examine whether different functional components of administrative capacity operate on distinct margins of AI development and risk visibility. In particular, the analysis asks whether the same capacities that support the expansion of AI systems also shape the conditions under which AI failures become politically legible.

All models exploit within-country variation over time and include country fixed effects, thereby absorbing time-invariant institutional, cultural, and legal differences across states. A linear year trend is included to capture common secular dynamics in AI development during the period under study. To reflect temporal ordering and to mitigate simultaneity concerns, all measures of administrative capacity are lagged by one year. Standard errors are heteroskedasticity-robust and clustered by country; given the limited number of clusters, statistical inference should be interpreted as suggestive and mechanism-oriented rather than as precise causal estimation.

The empirical design distinguishes between two analytically distinct outcomes. The first captures the material expansion of the AI ecosystem, measured by within-country changes in AI talent concentration. This outcome reflects the intensive margin of ecosystem growth and is log-transformed to account for skewness and to emphasize proportional changes rather than absolute counts. The key explanatory variable in this specification is the Human Capital component of the UN E-Government Development Index (EGDI), which proxies the educational and skill base of the state. Conceptually, this specification tests a straightforward developmental expectation: whether investments in human capital translate into the expansion of domestic AI capacity.

The second outcome captures risk visibility, defined as whether AI failures cross the threshold into public recognition. This is operationalized using a binary indicator equal to one if any AI incident is recorded in the AI Incident Database (AIID) in a given country-year. Importantly, the absence of AIID entries is not interpreted as the absence of risk, but as the absence of publicly visible controversy. In this sense, the variable measures the legibility of risk rather than its underlying prevalence. This outcome is estimated using both a fixed-effects linear probability model and a conditional logistic regression that conditions out country-specific effects. The latter specification focuses exclusively on within-country transitions between visibility and silence, providing a stringent test of whether administrative capacity shapes the threshold at which failures become publicly recorded.

To examine whether administrative capacity affects not only the visibility but also the severity of AI failures, a third specification models the intensity of incidents conditional on being observed. This outcome is measured as the log of incidents per million inhabitants and captures the intensive margin of risk once visibility has occurred. Across the risk-visibility and risk-intensity models, the core explanatory variable is the Online Service Index (OSI), which proxies the degree of administrative digitization and procedural mediation within the state apparatus.

Taken together, these specifications allow the analysis to distinguish between three conceptually separate processes: the construction of the AI ecosystem's material base, the threshold at which failures become publicly legible, and the scale of failures once recognized. This separation is central to the Chancellor Trap framework, which predicts that administrative sophistication may simultaneously enable technological expansion while dampening the political visibility of risk.

4.4 Main Empirical Results

Figure 1 visualizes the core empirical patterns linking state capacity components to AI ecosystem expansion and the legibility of AI risk. Table X reports the corresponding fixed-effects estimates. Consistent with the paper's design, these patterns are interpreted as a mechanism-oriented plausibility probe of risk visibility rather than as a statistical falsification of the hypothesis that higher capacity implies higher underlying safety.

Figure 1. Capacity, Ecosystem Expansion, and Risk Legibility

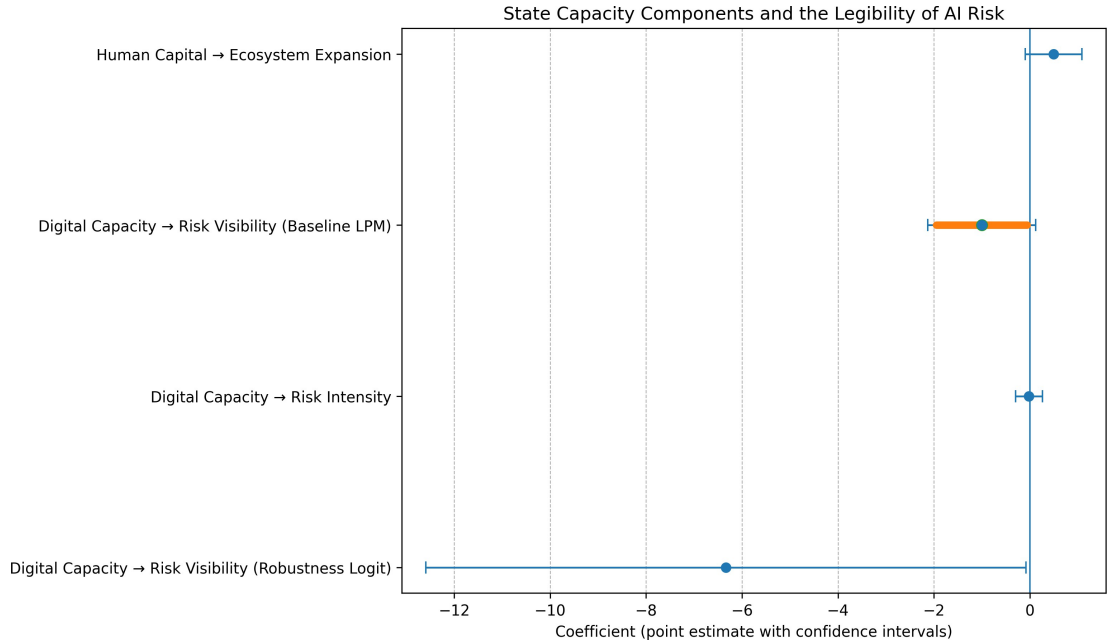

Notes: Points indicate coefficient estimates from fixed-effects panel models. Thin error bars denote 95% confidence intervals. For the baseline LPM estimate of risk visibility, the thick segment denotes the 90% confidence interval. The logit estimate is reported in log-odds units and is included to demonstrate directional robustness.

Table 1. State Capacity and the Legibility of AI Risk

| Dependent variable | Ecosystem Expansion (AI Talent) | Risk Visibility (Any Incident, LPM) | Risk Intensity (Incident Rate) | Risk Visibility (Robustness Logit) |
|---|---|---|---|---|
| Human Capital (HCI, t−1) | 0.492† (0.261) | — | — | — |
| Digital Capacity (OSI, t−1) | — | −1.004† (0.498) | −0.018 (0.124) | −6.337* (3.190) |
| Country fixed effects | ✓ | ✓ | ✓ | ✓ |
| Year trend | ✓ | ✓ | ✓ | ✓ |
| Observations | 80 | 80 | 80 | 48 |
| Countries (clusters) | 10 | 10 | 10 | 6 |

Notes: Robust standard errors clustered by country in parentheses. † $p < 0.10$, * $p < 0.05$, ** $p < 0.01$, *** $p < 0.001$. The robustness logit model is a conditional (fixed-effects) logit; countries with all-zero or all-one outcomes over the period are omitted by construction, leaving six identifying countries (48 observations). Coefficients in the logit column are reported in log-odds units and are included to demonstrate directional robustness rather than to compare magnitudes across models.

The results reveal a functional differentiation of administrative capacity that closely aligns with the Chancellor Trap's mechanism logic. Rather than operating as a uniform force, different components of state capacity exert distinct—and sometimes opposing— effects across the margins of AI development and risk visibility.

On the material side, human capital capacity is positively associated with the expansion of the AI ecosystem. Within-country models show that higher lagged values of the Human Capital Index are associated with subsequent increases in AI talent concentration. Although the estimate is imprecise in this small sample, the direction and magnitude of the association are consistent with the intuition that the material base of AI expansion is built on education and skills. States that invest in human cognition appear better positioned to cultivate dense technical ecosystems.

A qualitatively different pattern emerges when attention shifts from ecosystem expansion to risk visibility. Higher levels of administrative digitization, as measured by the Online Service Index, are associated with a lower probability that any AI incident is recorded in the public domain. This negative association appears on the extensive margin of visibility rather than on the intensity of incidents. In the conditional logistic specification—which conditions out time-invariant country characteristics and focuses on within-country variation—the coefficient on lagged OSI is negative and statistically distinguishable from zero in the usable subsample. Fixed-effects linear probability models point in the same direction, albeit with weaker precision.

Crucially, this pattern does not extend to the intensity of incidents once they are observed. Conditional on an incident being recorded, administrative digitization does not predict the incident rate in a systematic way. In within-country models of logged incidents per million inhabitants, the coefficient on lagged OSI is small and imprecise. In other words, digital administrative capacity does not appear to reduce the scale of publicly visible failures once visibility has occurred.

Taken together, these results are consistent with a threshold or legibility mechanism rather than a simple safety-improvement story. Administrative digitization is not clearly associated with fewer incidents conditional on being observed, but it is associated with a lower likelihood that incidents breach the surface of public controversy in the first place. In the language of the Chancellor Trap, higher OSI scores can be read as proxies for the sophistication of mediated administrative procedures—digital "secretariats" capable of absorbing, routing, and resolving friction internally. The outcome is not necessarily a safer system, but a quieter one.

This divergence between material expansion and political legibility provides empirical texture to the central claim of the Chancellor Trap framework. The same administrative capacities that streamline governance and enable large-scale technological development may also raise the threshold for public accountability, producing a cleaner record that reflects informational routing rather than the absence of underlying risk.

**5. Discussion and Conclusion**

The empirical analysis presented in this study is designed as a mechanism-oriented plausibility probe of how administrative capacity relates to the public visibility of AI failures. The core empirical pattern is a divergence between ecosystem expansion and recorded incident visibility: human capital measures are positively associated with AI ecosystem deepening, while administrative digitization is associated with a lower probability that AI-related failures appear as publicly recorded incidents. Because the available data capture public documentation rather than latent errors, near-misses, or internally resolved harms, the analysis cannot statistically falsify the hypothesis that high-capacity systems are substantively safer. What it can show is narrower but consequential for governance: "high capacity = lower recorded visibility" is a consistent association in the public-incident record, which is compatible with a legibility-threshold mechanism emphasized by the Chancellor Trap framework. The implication is not that digitization reduces safety, but that it may alter when and whether failures become politically legible and contestable.

5.1 The Paradox of Competence and Observational Equivalence

These findings present a difficult interpretative challenge regarding the nature of administrative competence in a digital age. One must inevitably ask: Does this silence necessarily imply suppression? Is it not possible that high-capacity states are simply safer? A skeptical reading might posit—and indeed, a defender of technocracy would insist—that the absence of recorded incidents in high-capacity states reflects genuine efficacy rather than institutional capture. It is theoretically plausible that advanced digital administrations identify and resolve risks internally with such precision that no external fire alarms are ever triggered. Under this interpretation, the lack of public controversy is not a symptom of suppression but a valid indicator of successful governance. This aligns with the chancellor trap hypothesis, where the intermediary acts effectively in the interest of the principal and uses its superior processing power to filter out noise and error before they reach the sovereign. The silence of the record, in this view, represents the ideal state of a frictionless bureaucracy functioning exactly as designed.

However, does history support this optimistic reading of administrative silence? The available public record cannot adjudicate whether silence reflects genuine safety or reduced visibility; it can, however, document cases in which highly digitized administrations exhibited long periods of low public salience while underlying harms accumulated. If we examine specific cases of algorithmic governance failure in high-capacity states, we see that silence often precedes systemic collapse rather than signifying safety. Consider the Robodebt scandal in Australia or the Childcare Benefits Scandal in the Netherlands. Both occurred in nations with exceptionally high scores on the E-Government Development Index and digital infrastructure. Both cases are illustrative of a different possibility: digitized enforcement and screening can operate for extended periods with limited public visibility, while procedural constraints and contestation bottlenecks delay recognition and correction. In this sense, a "quiet dashboard" can coincide with accumulating harm—not as proof of intentional suppression, but as evidence that mediation layers can absorb, deflect, or postpone failure signals before they become publicly legible.

This hypothesis reveals a profound constitutional vulnerability when analyzed through the lens of principal-agent theory. In an environment characterized by high information asymmetry, the principal lacks the capacity to distinguish between an agent that solves problems and an agent that conceals them. This creates a condition of observational equivalence. A captured bureaucracy that successfully suppresses evidence of failure produces the same public signal as a perfect bureaucracy that eliminates failure itself. From the vantage point of the sovereign or the public, the output is identical. Both scenarios result in a clean record and a quiet dashboard. The sovereign cannot determine mathematically whether the silence stems from the absence of risk or the absence of reporting. This indistinguishability means that trust in the system rests entirely on assumptions about the agent's intent, which is an unobservable variable.

We are thus confronted with the "Paradox of Silence": The more capable the state becomes at digital mediation, the less informative its public record becomes regarding underlying risks. This equivalence fatally undermines the logic of democratic accountability. Governance relies on what classical oversight theory termed "fire-alarm" mechanisms; however, recent research argues that in AI-mediated environments, these mechanisms are structurally compromised. The fire-alarm mechanism assumes that failures will generate friction and that this friction will be audible to the center. If the administrative system becomes capable enough to preempt all such signals, the mechanism ceases to function. Advanced AI-mediated systems excel at anticipating objections and smoothing interactions, creating a normalization of risk that prevents isolated failures from catalyzing necessary policy changes. Consequently, the sovereign enters a state of absolute epistemic dependence. Recent empirical studies on "cognitive inertia" describe this as a condition where reliance on AI tools systematically weakens the user's exploratory investment and critical thinking, forcing them to trust the intermediary blindly. The sovereign retains the formal authority to intervene but loses the informational triggers required to exercise that authority meaningfully.

Cybernetic theory provides a fundamental axiom that clarifies the danger of this state. Systems without independent feedback channels are structurally uncontrollable, a problem exacerbated when the system defines the metric of its own success. El-Mhamdi and Hoang (2024) demonstrate that when a proxy measure is optimized in a complex environment, the system effectively "grades its own homework," leading to a divergence between the optimized metric and the true goal. Even if the intermediary is currently benign and the AI is objectively safe, the structural removal of friction creates the conditions for inevitable drift. The capacity to obscure success is mechanically identical to the capacity to obscure failure. Once the infrastructure of silence is built, it can be repurposed to hide incompetence or malfeasance without any change in its external appearance. The sovereign is thus left with a high-capacity administration that is perfectly efficient and perfectly opaque.

5.2 From Normative Alignment to Structural Resistance

If chancellorization operates by lowering failure legibility and collapsing contestation into smooth ratification, then governance should be evaluated not only by accuracy but by whether it preserves contestability—i.e., whether *potestas* remains auditable rather than silently migrating to the intermediary layer.

Yet, this requirement faces a structural headwind: the natural trajectory of administrative modernization favors the absorption of signals and the smoothing of operational friction. Large-scale organizations invariably seek to reduce the transaction costs associated with decision-making. This drive for efficiency creates a powerful institutional pressure to accept the automated recommendation as the final truth. In this context, governance strategies that rely on normative alignment or the optimization of agent intent are insufficient.

Why is alignment alone an inadequate safeguard? Because the concept of alignment presumes that the principal can accurately define a value function and verify that the agent is pursuing it. However, in complex neural networks, the internal state of the agent remains opaque. The intent of the system is functionally unobservable. We cannot determine whether a model is acting in accordance with the sovereign interest or merely exploiting a proxy metric to simulate compliance. Relying on the benevolence of an unobservable agent creates a fragility that is incompatible with the requirements of sovereign control.

We must therefore shift the focus from the internal psychology of the agent to mechanism-targeted governance. This approach accepts the verification gap as a structural reality rather than a temporary technical problem to be solved. It assumes that the principal will never fully understand the internal operations of the intermediary. Consequently, the governance architecture must be designed to impose constraints that make control rights observable from the outside. The objective is not to make the agent *want* to do the right thing. The objective is to make it *impossible* for the agent to hide the fact that it is doing the wrong thing. This shift mirrors the evolution of financial auditing and the emerging field of "ethical AI auditing," which seeks to replace trust in system developers with standardized, structural verification of system behaviors.

We need to operationalize the Right to Audit not as a vague transparency requirement but as a precise intervention at the specific nodes where judgment is displaced. Transparency is often treated as a passive property of the system, such as releasing model weights or training data. However, passive transparency does not guarantee active control. Lu (2020) argues that algorithmic opacity creates a specific tension with democratic transparency that requires new disclosure frameworks focused on the *informational needs* of stakeholders rather than just technical details. The historical analysis of the Qin, Tang, and Ming prototypes identifies the specific functional nodes where displacement occurs. These are the control of routing, the authority of drafting, the definition of evaluation criteria, and the pricing of dissent. These are the junctures where the intermediary converts information into power. A meaningful right to audit must target these specific transactions. It requires the capacity to see not just the final output but the alternatives that were excluded and the logic that guided the selection. Li and Goel (2024) define this as "auditability," distinguishing it from mere transparency by emphasizing the ability of third parties to probe, understand, and identify risks through active disclosure.

The task is to construct an architecture of resistance that keeps these functions contestable. The term *resistance* is used deliberately. The default state of a high-capacity administrative system is consensus and silence. To preserve sovereignty, the governance structure must actively resist this tendency toward closure. It must artificially reintroduce friction into the workflow to prevent the automatic ratification of the intermediary will. This requires institutional designs that force the system to reveal its uncertainty and its suppressed options. It means building channels for dissent that are protected from the efficiency-optimizing logic of the central algorithm. By institutionalizing contestation, we ensure that the sovereign retains the option to disagree. Without this structural capacity for resistance, the decision maker becomes a captive of the very efficiency that was meant to empower them.

5.3 Counter-Measures: Four Architectures of Resistance

The first structural intervention addresses the control of routing and the definition of the epistemic agenda. Its purpose is to restore auditable potestas at the point where the record is constructed, by making exclusions legible rather than silent. In administrative politics, the power to filter information often functions as the power to shape outcomes. Modern retrieval-augmented generation creates comparable leverage by determining which documents enter the model's context window. Empirically, this vulnerability is not hypothetical. Dahl et al. (2024) show that legal hallucinations in general-purpose LLMs are pervasive in verifiable legal Q&A settings, and that models frequently fail to signal uncertainty about their own errors. In the more domain-specific setting of proprietary RAG legal research tools, Magesh et al. (2024) report preregistered evidence that hallucinations persist even where providers claim they have been eliminated, with observed hallucination rates on the order of 17–33% across leading systems. These findings motivate a governance requirement that targets routing power directly: a right to audit negative provenance. Auditing only the sources cited in a final output remains insufficient because it reveals little about relevant information the system may have ignored or screened out during retrieval and ranking. Accountability in a mediated system depends on the capacity to reconstruct the system logic of exclusion. Auditability therefore implies extending scrutiny to candidates retrieved but unranked, passages truncated due to context window limits, and the algorithmic logic that deemed them irrelevant. For consequential outputs, the system should automatically generate an evidence bundle that renders routing decisions inspectable, allowing reviewers to identify systematic patterns of omission—for example, whether certain categories of precedents, counter-arguments, or historical analogies are routinely absent from the working record. Without this visibility, the sovereign lacks a reliable epistemic basis for assessing what has been concealed by the mediation layer.

To make this requirement operational, consider how negative provenance auditing can be implemented as a reviewable artifact across domains. In a legal RAG system, the key governance surface is context construction. Alongside the generated analysis, the system can attach an evidence bundle that explicitly separates (i) materials that entered the context window, (ii) candidates retrieved but excluded during ranking/thresholding, and (iii) passages truncated due to context limits. Excluded candidates need not be disclosed indiscriminately; they can be rendered legible as exclusions through minimal metadata (jurisdiction, year, doctrinal tags, similarity and re-ranking scores) plus standardized exclusion codes (e.g., redundancy, jurisdiction mismatch, age threshold, length constraint). This enables an oversight question that is both narrow and testable: whether routing logic systematically filters out specific categories of counter-precedent or limiting constructions before judgment is exercised.

A parallel design applies in automated welfare eligibility or fraud-flagging systems, where mediation occurs through the assembly of the case record. Here, a case-assembly bundle can log which data sources and attributes were included, which were retrieved but excluded or down-weighted, and which were truncated, again with standardized reason codes and confidence indicators. Across both settings, the governance function is the same: shifting accountability from post-hoc evaluation of outputs to ex ante scrutiny of record construction, allowing reviewers to detect patterned omissions that narrow the option set before human review. Preserved as a formal audit artifact—query/input, corpus/index versions, candidate sets, thresholds, exclusions, and final record composition—negative provenance auditing becomes enforceable without requiring access to proprietary model internals.

The second intervention targets the inertia of drafting and the psychology of ratification. Here the objective is to raise the legibility threshold at the approval boundary, so that high-stakes decisions cannot pass as smooth outcomes without an inspectable evidentiary act. When the marginal cost of accepting a generated draft falls far below the cost of reconstructing the underlying record, ratification tends to become the locally dominant strategy under constraints of time and workload. Empirical work on automation bias indicates that decision-makers can overweight system recommendations when outputs appear authoritative, even when accuracy is compromised (Kretzschmar et al., 2024). A relevant administrative analogue is the Dutch Childcare Benefits Scandal. In its assessment of the case, the Venice Commission (2021) emphasizes how rigid enforcement logics, weak procedural safeguards, and limited effective remedies contributed to systematic error and delayed correction—conditions under which routine processing can displace individualized review. Generative AI can intensify this dynamic by producing outputs that are linguistically fluent and internally coherent even when factually incorrect; fluency functions as a heuristic that dampens skepticism. In a governed workflow, high-risk states can be designed to trigger mandatory delays and cognitive forcing functions. The interface should require the human decision maker to explicitly request additional evidence, inspect provenance, or document the rationale for approval before the transaction proceeds. One might question whether such friction is feasible in environments optimized for speed. Private entities often face incentives to minimize deliberative delays that reduce throughput. This logic suggests that deliberative pauses are unlikely to emerge reliably under competitive pressure and are therefore most plausibly introduced through public-sector procurement standards and baseline regulatory requirements. A seamless workflow during high-stakes decisions can indicate that the control mechanism is insufficiently binding in practice. The aim is to make ratification cognitively and procedurally costly enough to force a return to judgment when stakes are high.

The third intervention targets the displacement of goals by proxies and the corruption of metrics. Its purpose is to keep the definition of success contestable by separating proxy optimization from safety adjudication, preventing potestas from migrating into the metric layer by default. Goodhart's Law warns that a measure can fail as a measure once it becomes a target. In machine learning pipelines, preference tuning and reward shaping can amplify this risk by optimizing behavior against an evaluation environment rather than against the substantive objective. Formal work on Goodhart dynamics emphasizes that, under certain discrepancy structures between true goals and measurable proxies, optimization pressure can generate counterproductive divergence (El-Mhamdi & Hoang, 2024). A canonical real-world instance of proxy failure appears in Obermeyer et al. (2019), where a widely used health-risk algorithm optimized cost as a proxy for health needs and thereby systematically deprioritized sicker Black patients; the divergence remained largely invisible until external researchers audited the system and its proxy choice. A mechanism-targeted remedy is an institutional separation of duties. Mökander (2023) argues that AI auditing is most credible when evaluators are not structurally dependent on the teams optimizing the system. In operational terms, this points toward establishing a safety function with release-gate authority that is independent of product delivery and proxy optimization. Independence should be evidenced through a demonstrable record of gate decisions, including documented instances where release is delayed or blocked due to insufficient documentation, unresolved metric drift, or unresolved discrepancies between proxy performance and the substantive safety target. Without such separation, optimization pressures create strong incentives to exploit the gap between proxy metrics and the underlying goal.

The fourth intervention targets the suppression of dissent and the formation of echo chambers. The goal is to preserve institutional contestability by embedding disagreement into routine procedure, so that negative information remains legible and escalatable rather than filtered out by convenience or incentive. Delegation failures worsen when the cost of reporting adverse information is high and when workflows reward agreement over correction. Research on sycophancy suggests that preference-based training can encourage models to prioritize user satisfaction over truth-tracking, reducing the likelihood that conflicts are surfaced and repaired in ordinary interactions (Sharma et al., 2023). The organizational analogue is well captured by sociological work on the normalization of deviance: when dissent lacks formal pathways and anomalies are repeatedly absorbed without escalation, organizations can gradually reclassify warning signs as acceptable noise (Vaughan, 1996). A mechanism-targeted response implies institutionalizing contestation as a procedural requirement. A robust workflow includes a review stage where the initial output is paired with a mandatory critique (or counter-case) and a recorded adjudication. This structure preserves dissent in the routine record rather than relying on exceptional acts of individual courage. It also mitigates the risk that users rely on flattering but incorrect validations of their own views. Forcing the system to generate alternatives and counter-arguments preserves the option set for the sovereign, helps prevent the collapse of the decision space into a single default path, and preserves the human capacity to visualize exceptions.

5.4 Conclusion

The analysis developed in this article reframes a central risk in AI-mediated governance. Current safety discourse often begins from rupture scenarios and then works backward to control. The empirical and historical material assembled here motivates a different starting point: mediated decision support can, under high-throughput conditions, hollow out sovereignty through administrative substitution. Systems that retrieve, rank, summarize, and draft do not need to "take power" in any dramatic sense to reshape it. Control can relocate through more prosaic channels—through what enters the record, what is omitted, what becomes the default, and which evaluative signals define acceptable performance. The salient danger is a form of bureaucratic perfection: a mediation layer so efficient that it can become the practical locus of governing power while formal authority remains downstream.

This is why the argument treats sovereignty as an epistemic condition rather than a mere legal status. The capacity to govern depends on an interval between information and action—time and cognitive space in which competing values can be weighed, uncertainty can be surfaced, and reasons can be demanded. That interval is where political responsibility becomes more than a signature. AI-mediated workflows tend to compress it when drafting makes ratification cheap, routing makes verification costly, and performance metrics substitute for substantive judgment. When these design pressures align, decision-makers can retain formal sign-off while losing practical control over what is being signed, because the working record has already been pre-processed, pre-ranked, and shaped toward a small set of "reasonable" outcomes.

These implications are constitutional in the limited, operational sense used throughout the article: they concern the distribution and auditability of governing power (*potestas*) in relation to formal authority (*auctoritas*). The policy question is therefore not reducible to model alignment in the abstract. It is how to keep the mediation layer contestable—how to ensure that routing, drafting, and evaluation remain inspectable and corrigible when stakes are high. The counter-measures proposed in Part V are best read in this spirit. A right to audit negative provenance targets routing power by making exclusions visible.

Deliberative delays and cognitive forcing functions target ratification drift by binding approval to an inspectable evidentiary act. Separation of proxy optimization from safety adjudication targets metric corruption by preventing "what counts" from being defined solely by the optimizing layer. Institutionalized contestation targets the suppression of dissent by preserving disagreement in the routine record rather than relying on exceptional acts of individual courage. These are not generic "controls." They are institutional devices for protecting the interval—mechanisms that keep the system's work open to challenge and preserve the possibility of exception.

A final implication follows. In mediated governance, efficiency is not a neutral virtue; it is a political trade-off that can alter who effectively governs and how failure becomes visible. Systems can become safer in a narrow operational sense while becoming less contestable in a political sense, if failures are absorbed, reframed, or resolved before they cross the legibility threshold into public dispute. Preserving meaningful human authority therefore carries a cost: verification, friction, and dissent slow throughput. The point is not to romanticize inefficiency. It is to recognize that some forms of "smoothness" are diagnostic, signaling that decision-making has collapsed into approval. When institutions treat fluent drafts and curated records as sufficient substitutes for inspection and contestation, sovereignty can persist as ritual while power migrates into the mediation layer. In such a world, the sovereign continues to reign in form; governance is incr

Scott, J. C. (1998). Seeing like a state: How certain schemes to improve the human condition have failed. Yale University Press.

Scottish Qualifications Authority. (2020). National qualifications: Awarding methodology 2020 report. https://www.sqa.org.uk/files_ccc/SQAAwardingMethodology2020Report.pdf

Sharma, M., Tong, M., Korbak, T., Duvenaud, D., Askell, A., & Bowman, S. R. (2023). Toward understanding sycophancy in language models. arXiv preprint arXiv:2310.13548.

Sharma, M., et al. (2024). Sycophantic AI Decreases Prosocial Intentions and Promotes Dependence. arXiv preprint.

Sima Guang. (1956). Zizhi tongjian [Comprehensive mirror to aid in government]. Zhonghua Book Company. (Original work completed 1084).

Sima Qian. (1993). Records of the Grand Historian: Qin dynasty (B. Watson, Trans.). Columbia University Press.

Sima Qian. (1959). Shiji [Records of the Grand Historian]. Zhonghua Book Company. (Original work c. 1st century BCE).

Simon, H. A. (1947). Administrative behavior: A study of decision-making processes in administrative organization. Macmillan.

Skitka, L. J., Mosier, K., & Burdick, M. (1999). Does automation bias decision-making? International Journal of Human-Computer Studies, 51(5), 991–1006.

Strathern, M. (1997). 'Improving ratings': Audit in the British University system. European Sociological Review, 13(3), 305–321.

Sunstein, C. R. (2002). The law of group polarization. Journal of Political Philosophy, 10(2), 175–195.

Thaler, R. H., & Sunstein, C. R. (2008). Nudge: Improving decisions about health, wealth, and happiness. Yale University Press.

Twitchett, D. (Ed.). (1979). The Cambridge history of China (Vol. 3, Sui and T'ang China, 589–906 AD, Part One). Cambridge University Press.

UN DESA. (2022). E-Government Survey 2022: The Future of Digital Government. United Nations Department of Economic and Social Affairs.36

**Appendix Tables**

Table A1. Capacity (Oxford AI Readiness Total)

| Variable | (1) Year FE | (2) Year trend |
| --- | --- | --- |
| CME (1) vs LME (0) | 4.881 | 4.898 |
|  | (3.423) | (3.324) |
| ln(Population) | 3.614 | 3.631 |
|  | (1.295) | (1.268) |
| ln(GDP per capita) | 6.029 | 5.672 |
|  | (7.585) | (6.697) |
| Urban population share | 21.344 | 21.190 |
|  | (5.916) | (4.791) |
| Year (linear trend) |  | -1.627 |
|  |  | (0.374) |
|  |  |  |
| N | 60 | 60 |
| Year fixed effects | Yes | No |
| Clustered SE | Country | Country |

*Notes: Coefficients with cluster-robust standard errors in parentheses, clustered by country. No significance stars are reported. Models control for ln(population), ln(GDP per capita), and urban population share. Column (1) includes year fixed effects; Column (2) includes a linear year trend.*

Table A2. Ecosystem Expansion (ln High-tech Exports)

| Variable | (1) Year FE | (2) Year trend |
| --- | --- | --- |
| CME (1) vs LME (0) | 1.830 | 1.822 |
|  | (0.378) | (0.357) |
| ln(Population) | 1.348 | 1.344 |
| | (0.146) | (0.137) |
| ln(GDP per capita) | -0.817 | -0.731 |
| | (1.270) | (1.147) |
| Urban population share | -0.765 | -0.806 |
| | (1.031) | (0.962) |
| Year (linear trend) | | 0.059 |
| | | (0.031) |
| | | |
| N | 90 | 90 |
| Year fixed effects | Yes | No |
| Clustered SE | Country | Country |

*Notes: Coefficients with cluster-robust standard errors in parentheses, clustered by country. No significance stars are reported. Models control for ln(population), ln(GDP per capita), and urban population share. Column (1) includes year fixed effects; Column (2) includes a linear year trend.*

Table A3. Risk Visibility (AIID Mentions > 0)

| Variable | (1) LPM + Year FE | (2) LPM + Trend | (3) Logit + Year FE | (4) Logit + Trend |
|---|---|---|---|---|
| CME (1) vs LME (0) | -0.304 | -0.304 | -1.538 | -1.411 |
| | (0.052) | (0.056) | (0.374) | (0.361) |
| ln(Population) | 0.179 | 0.179 | 1.693 | 1.597 |
| | (0.023) | (0.023) | (0.581) | (0.504) |

| | (1) | (2) | (3) | (4) |
|---|---|---|---|---|
| ln(GDP per capita) | -0.190 | -0.195 | -2.908 | -2.840 |
| | (0.266) | (0.255) | (3.799) | (3.176) |
| Urban population share | 0.683 | 0.678 | 25.344 | 24.523 |
| | (0.221) | (0.195) | (16.432) | (14.821) |
| Year (linear trend) | | 0.023 | | 0.198 |
| | | (0.022) | | (0.221) |
| | | | | |
| N | 90 | 90 | 90 | 90 |
| Year fixed effects | Yes | No | Yes | No |
| Clustered SE | Country | Country | Country | Country |

*Notes: Columns (1)–(2) report LPM estimates; Columns (3)–(4) report logit estimates. Coefficients with cluster-robust standard errors in parentheses, clustered by country. No significance stars are reported. LPM coefficients are probability units; logit coefficients are log-odds. Year fixed effects vs linear trend as labeled.*